# Large low-temperature magnetoresistance in SrFe$_2$As$_2$ single crystals

S. V. Chong[1(a)], G. V. M. Williams[2], J. Kennedy[3], F. Fang[3], J. L. Tallon[1] and K. Kadowaki[4]

[1] Callaghan Innovation Research Limited - P.O. Box 31310, Lower Hutt 5040, New Zealand
[2] MacDiarmid Institute, School of Chemical and Physical Sciences, Victoria University of Wellington - P.O. Box 600, Wellington 6140, New Zealand
[3] National Isotope Centre, GNS Science - P.O. Box 31312, Lower Hutt 5040, New Zealand
[4] Institute of Materials Science and Graduate School of Pure & Applied Sciences, University of Tsukuba - 1-1-1, Tennodai, Tsukuba, Ibaraki 305-8573, Japan

**Abstract** – We present the first report on a large low-temperature magnetoresistance (MR) of more than 1600% in a SrFe$_2$As$_2$ single crystal and 1300% in a low-energy Ca ion-implanted SrFe$_2$As$_2$ single crystal that occurs before the emergence of crystallographic strain-induced bulk superconductivity arising from a sample aging effect. In accordance to band structure calculations from literature, which consitently show more than 2 bands are involved in the transport, we have modeled this large MR at high fields using a 3-carrier scenario rather than solely on quantum linear MR model generally used to explain the MR in iron-pnictides. At and below 20 K the large MR may be due to 3-carrier transport in an inhomogeneous state where there are superconducting and metallic regions.

**Introduction.** – There is considerable interest in the recently discovered iron-based superconductors because the superconducting transition temperature is as high as 56 K and superconductivity arises from antiferromagnetic or spin density wave (SDW) metallic parent compounds [1-4]. Furthermore, there are reports of the co-existence of superconductivity and magnetism particularly in the underdoped region [5,6]. Like the cuprates and heavy fermions, the iron-based pnictides and chalcogenides are unconventional superconductors, with the notable feature that a single-carrier model is inadequate to characterize their transport behavior due, for example, to the multiband nature of the $d$-orbitals involved in the conduction bands.

Ran *et al.* predicted that exotic Dirac cone states could occur at spin density wave nodes that arise from the topology and symmetry of the band structure [7]. A Dirac-cone-like dispersion was later observed in BaFe$_2$As$_2$ by angle resolved photoemission spectroscopy (ARPES) where the Fermi level was close to the cone apex and hence small Fermi surfaces were predicted [8]. Dirac cones are interesting because they should have nearly zero effective mass, $m^*$, and can have very high mobilities [8-13]. Furthermore, quantum linear magnetoresistances are predicted at high fields and when only one Landau level is occupied. In this case the resistivity can be written as $\rho_{xx} = \rho_{yy} = [N_i/(\pi n_D^2 e)] \times B$, where $N_i$ is the concentration of static scattering centers, $n_D$ is the density of carriers, and $B$ is the applied magnetic field [14]. It should be noted that the quantum linear regime assumes that the Fermi level is near the cone apex. However, if the Fermi level is significantly above or below the crossing point then more than one Landau level will be filled at high fields and hence quantum transport may be observed but the magnetoresistance (MR) will not necessarily be proportional to $B$.

A linear MR was observed in BaFe$_2$As$_2$ and attributed to Dirac Fermions [10,11]. However, not all BaFe$_2$As$_2$ studies show linear MR behaviour [12,15]. For example, it was reported that a linear MR occurs in as-made BaFe$_2$As$_2$ samples and annealing removed lattice distortions and defects resulting in a non-linear MR [12]. The MR was described using a 3-carrier model and magnetoresistances of ~280% were observed at 5 K and 7 T. Terashima *et al.* concluded from a Shubnikov de Haas study of BaFe$_2$As$_2$ that the magnetic fields used in the MR studies (<10 T) are far too low for any quantum linear transport to be observed [15]. Quantum oscillation measurements on SrFe$_2$As$_2$ were interpreted in terms of a Dirac cone dispersion near the Fermi level [16]. However, in this case the MR does not appear to be linear in the applied magnetic field. Most MR reports have focused on non-superconducting BaFe$_2$As$_2$ and hence there is a need to study SrFe$_2$As$_2$.

Here we report the results from MR measurements on SrFe$_2$As$_2$ single crystals. MR measurements were also done on a low-energy Ca ion-implanted SrFe$_2$As$_2$ single crystal with an average Ca concentration of 0.8%, where this

(a) E-mail: s.chong@irl.cri.nz

concentration was chosen to ensure that superconductivity is not induced by Ca ion implantation that is known to occur for 6% Ca [17]. The 0.8% Ca-implanted crystal was selected for this study partly because we wanted to see if this leads to a near-surface contribution to the MR. It was also selected because we wanted to see if the MR mechanism is similar for different crystals. We show below that there is a huge increase in the low-temperature MR when the crystals are aging and before the appearance of bulk strain-induced superconductivity. Furthermore, the MR can be modeled by 3-carrier transport and the mobilities and carrier concentrations are not significantly enhanced by low-energy Ca implantation.

**Experimental details.** – The $SrFe_2As_2$ single crystals were taken from the same batch used in our previous study [17]. They were initially annealed in ultra-pure argon to remove any trace of superconductivity using the process described by Saha *et al.* [18] and then placed in a vacuum desiccator with a pressure of ~0.2 bar. Low-energy calcium ion implantation was carried out at the GNS Science ion implantation facility using a 20 keV beam energy, similar to our previous work [17]. An estimated implantation depth of 10 to 15 nm was determined from Monte Carlo simulation [19]. The fluence of $1 \times 10^{15}$ ions/cm$^2$ corresponds to an average Ca concentration in the implanted region of 0.8 atomic %. Resistivity and MR measurements were carried out using the four-terminal configuration on a *Quantum Design* Physical Property Measurement System (PPMS) up to ± 8 Tesla. The current was perpendicular to the applied magnetic field and the magnetic field was parallel to the c-axis.

**Results and analysis.** – Figure 1 shows the resistivity of a post argon annealed unimplanted $SrFe_2As_2$ single crystal with a residual resistivity at 4 K of 30 μΩ·cm (curve (a)), which indicates good sample quality. Below the SDW magneto-structural transition at 210 K for this compound, the temperature dependence of the resistivity, $\rho$, follows a metallic $T^2$ behavior, that is $\rho(T)=\rho_0+aT^2$, where $a$ is a constant and referred as the Fermi-liquid coefficient [20], $T$ is the temperature, and $\rho_0$ is the residual resistivity. After the sample was left in the desiccator for more than six months (fig. 1, curve (b)), the resistivity increased and showed a downturn at 25 K. However, the residual resistivity did not fall to zero even at 2 K. Similar $T$-dependent behavior was also observed in the 0.8% Ca as-implanted $SrFe_2As_2$ (fig. 1(d)) left in the vacuum desiccator for 42 days. Unlike the 6% Ca-implanted crystal [17], the resistivity is still finite even at the lowest temperature measured. A downturn in the resistivity at low temperatures has also been reported from pressure-dependent measurements on $AFe_2As_2$ ($A$ = Ba, Sr or Ca) where the crystals were fully superconducting at high-pressure [21-23]. The initial low-temperature downturn in the resistivity indicates the presence of an inhomogeneous electronic state where there are superconducting and non-superconducting regions, and where the superconducting fraction is below the percolation threshold.

After an additional 3 months in the desiccator, both the unimplanted and Ca-implanted crystals became fully superconducting with a superconducting transition temperature, $T_c$, ~24 K (fig. 1, curve (c) and (e)) similar in value to $T_c$ that we previously reported from measurements on ion-implanted $SrFe_2As_2$ superconductors [17]. The appearance of superconductivity is most probably due to a strain-induced superconductivity that occurs when $SrFe_2As_2$ crystals are exposed to air [18], where this process is reversible by annealing in argon. Water is likely to play a role in this process because it is known that exposure to water vapor leads to a superconducting phase [24]. Furthermore, similar to our previous study we find that the SDW transition remains fixed at 210 K, which suggests that charge-doping is unlikely to be involved in inducing superconductivity [18].

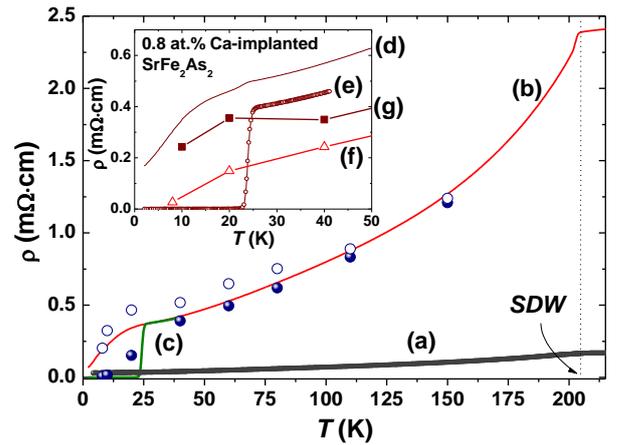

Fig. 1: (Colour on-line) Temperature dependence of the resistivity of $SrFe_2As_2$ (a) after annealing, (b) after 6 months, and (c) after 9 months in a vacuum desiccator. The inset shows the resistivity data for a 0.8% Ca-implanted $SrFe_2As_2$ crystal (d) as-implanted after 42 days, and (e) stored for 3 months in the same vacuum desiccator as the non-implanted $SrFe_2As_2$ crystal. Also shown is the resistivity measured after 7 months at 0 T (filled circles) and 8 T (open circles) for the unimplanted $SrFe_2As_2$ crystal. (f) and (g) are the resistivity calculated from the fitted parameters from fig. 2 and 3, respectively.

The MR of unimplanted $SrFe_2As_2$ was measured after 7 months where there was an additional decrease in the resistivity below the incipient $T_c$ (fig. 1, filled circles). Here we define the MR as $MR=(\rho(B)-\rho(0))/\rho(0)$, where $\rho(0)$ is the resistivity at zero field and $\rho(B)$ is the resistivity in an applied magnetic field. The resultant MR down to 40 K is plotted in fig. 2(a) as a function of the applied magnetic field. The MR is negligible above the SDW transition temperature at 210 K and increases rapidly below this temperature. Furthermore, the MR at 40 K reaches over 30% at 8 T, very similar to that observed in our previous work on ion-implanted $SrFe_2As_2$ single crystals [17]. It is larger than the 9.5% at 40 K and 9 T observed in a $SrFe_2As_2$ thin film [24]. A similarly high MR of 25% was previously reported by Chen *et al.* from MR measurements on non-superconducting SrFe$_2$As$_2$ single crystals at a lower temperature of 10 K and for an applied magnetic field of 8 T [25]. Recently, Ishida *et al.* reported very large MR values in BaFe$_2$As$_2$ single crystals co-annealed with BaAs powder, with MR~170% at 30 K and 7 T [12].

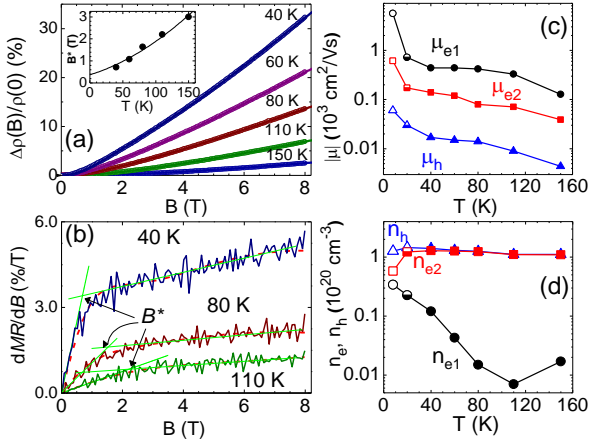

Fig. 2: (Colour on-line) (a) Magnetoresistance and (b) the derivative of the MR for SrFe$_2$As$_2$ left in a vacuum desiccator for 7 months. The solid curves in (a) are 3-carrier fits to the data and the derivatives at 40, 80 and 110 K are plotted in (b) (dashed curves). The solid lines in (b) show the high- and low-field extrapolations and the method used to estimate $B^*(T)$. $B^*(T)$ is plotted in the inset to (a). (c) The carrier mobilities and (d) concentrations obtained from 3-carrier fits to the data in Fig. 2(a) (solid circles symbols). Also shown are the mobilities and carrier concentrations from 3-carrier fits to the data in Fig. 4(a) at 8 K (open symbols), and 20 K (half-filled symbols).

The derivatives of the MR at 40, 80 and 110 K are plotted in fig. 2(b). It can be seen that there is a crossover in the MR derivatives and this crossover field, $B^*$, is plotted as a function of temperature in the inset of fig. 2(a). A crossover was reported from MR measurements on BaFe$_2$As$_2$ and Ba(Fe$_{1-x}$Ru$_x$As)$_2$ where the crossover was from $MR \propto B^2$ to $MR \propto B$ [10,11]. It was argued that the crossover was due to semiclassical 2-carrier transport for low fields and a Dirac cone with mass-less Dirac Fermions at high fields where the MR is proportional to $B$ in this quantum linear transport regime. The crossover field, $B^*$, has been defined as [10,11,14],

$$B^* = (1/2e\hbar v_F^2)(k_B T + E_F)^2 \qquad (1)$$

where $v_F$ is the Fermi velocity, $k_B$ is Boltzmann's constant, and $E_F$ is the Fermi energy. Equation (1) was derived by assuming that the crossover field occurs when $\Delta_1 = k_B T + E_F$, where $\Delta_1$ is the field-induced splitting of the first Landau level, and $\Delta_1 = v_F(2\hbar eB)^{1/2}$. In our case, although a crossover can clearly be seen in fig. 2(b), the MR is not linear at high fields and hence the high-field MR data cannot be completely attributed to quantum linear transport. However, despite this, we show that $B^*$ in the inset to fig. 2(a) can be fitted using eq. (1) where we find that $E_F = 6.9 \pm 1.7$ meV and $v_F = (3.11 \pm 0.26) \times 10^5$ ms$^{-1}$. $E_F$ is slightly higher than that reported from MR measurements on BaFe$_2$As$_2$ (2.48 meV, Tanabe *et al.* [11]) but it is within the range reported from MR measurements on Ba(Fe$_{1-x}$Ru$_x$)$_2$As$_2$ (3-12 meV, Huynh *et al.* [10]). $v_F$ is slightly higher compared to $v_F \sim 2 \times 10^5$ ms$^{-1}$ reported in both studies.

Our analysis of the MR data suggests that the MR at high fields could contain a contribution from Dirac Fermions as well a contribution from semiclassical multi-carrier transport from other bands that could account for the departure from the expected quantum linear transport. However, it is possible, and probably more likely, that the MR data in fig. 2(a) is dominated by multi-carrier transport, similar to that reported by Ishida *et al.* [12] from measurements on BaFe$_2$As$_2$. The authors attributed their MR to a change in the carrier mobility and concentration with decreasing temperature, where one carrier could possibly be attributed to carrier transport from a band with a Dirac-cone-like dispersion. We have also modeled our MR data using a multi-carrier model. We find that a 3-carrier model with one hole band and two electron bands is required to fit our data. This is reasonable given the band structure and the appearance of hole and electron pockets at the Fermi level where three types of carriers have recently been reported from quantum oscillation measurements on BaFe$_2$As$_2$ [15].

We used the 3-carrier matrix formalism for the (longitudinal) MR developed by Kim [26] to analyze our MR data below the SDW transition. In this case the MR can be written as,

$$MR = [(\alpha + \gamma B^2) B^2]/[1 + (\beta + \delta B^2) \times B^2] \qquad (2)$$

where $\alpha = f_1 f_2 (\mu_1 - \mu_2)^2 + f_1 f_3 (\mu_1 - \mu_3)^2 + f_2 f_3 (\mu_2 - \mu_3)^2$; $\gamma = f_1 f_2 (\mu_1 - \mu_2)^2 \mu_3^2 + f_1 f_3 (\mu_1 - \mu_3)^2 \mu_2^2 + f_2 f_3 (\mu_2 - \mu_3)^2 \mu_1^2$; $\beta = (f_1 \mu_2 + f_2 \mu_1)^2 + (f_3 \mu_1 + f_1 \mu_3)^2 + (f_2 \mu_3 + f_3 \mu_2)^2$; $\delta = (f_1 \mu_2 \mu_3 + f_2 \mu_1 \mu_3 + f_3 \mu_1 \mu_2)^2$; where $f_i = |n_i \mu_i|/\sum |n_i \mu_i|$, $n_i$ and $\mu_i$ are the carrier density and mobility, respectively. We assume the total carrier density is in the order of $10^{20}$ cm$^{-3}$ below the SDW transition in accordance to optical conductivity measurements [27], Hall measurements [28] and DFT calculations [29], which all show $n_i$ varies with temperature, and $n_h = n_{e1} + n_{e2}$, where $n_h$, $n_{e1}$, and $n_{e2}$ are the hole and two electron concentrations, respectively. Similar assumptions were also made by Ishida *et al.* for their fitting of the non-superconducting BaFe$_2$As$_2$ magnetic-field dependent Hall resistivity data [12]. The fit to our MR data is good over the measured magnetic field range (fig. 2(a), solid curves) where the carrier mobilities and concentrations are plotted in fig. 2(c) and fig. 2(d), respectively. We show in fig. 2(b) that the 3-carrier model also provides a good fit to the derivative of the MR (dashed curves). Moreover, the zero field resistivity obtained from the fitted parameters, $\rho_{xx}(0) = 1/\sum |en_i \mu_i|$, are in the same order of magnitude as the experimental data as shown in fig. 1(f) for SrFe$_2$As$_2$ and (g) for the Ca-implanted $SrFe_2As_2$ sample presented below. We have tried to vary $n_i$ to see if there is a significant change in $|\mu_i|$ or the temperature dependence behaviour by setting: (a) $n_e = n_h = 5 \times 10^{20}$ cm$^{-3}$; (b) $n_e = 2 \times n_h$; (c) $n_e = 10 \times n_h$; and (d) making $n_i$ temperature independent like Ishida *et al.*, using $n_i$ (40 K). The fits to the MR are reasonable and while the resultant magnitude of $\mu_i$ changes, we do not find any major changes in the temperature dependences and we still find that $\mu_{e1} > \mu_{e2} > \mu_h$. However, we find that the fitted $\mu_i$ and $n_i$ now underestimate the $\rho_{xx}(0)$ by a factor of 2 to 28. The $\mu_i$ and $n_i$ values we report herein gave a much consistent $\rho_{xx}(0)$ value.

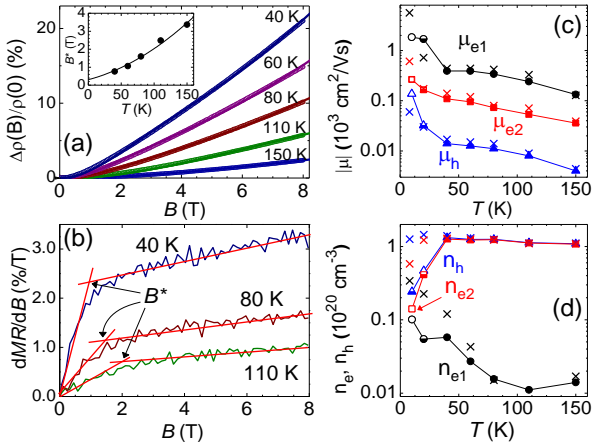

Fig. 3: (Colour on-line) (a) Magnetoresistance and (b) the derivative of the MR for the 0.8% Ca-implanted crystal left in a vacuum desiccator for 72 days. The solid curves in (a) are 3-carrier fits to the data. The solid lines in (b) show the high- and low-field extrapolations and the method used to estimate $B^*(T)$. $B^*(T)$ is plotted in the inset to (a). (c) The carrier mobilities and (d) concentrations obtained from 3-carrier fits to the data in Fig. 3(a) (solid circles, squares, and triangles). Also shown are the mobilities and carrier concentrations from 3-carrier fits to the data in Fig. 4(b) at 10 K (open symbols), and 20 K (half-filled symbols) as well as from 3-carrier fits to the MR data from the unimplanted $SrFe_2As_2$ crystal plotted in Fig. 2(a) and Fig. 4(a) (crosses).

Our results are similar to those reported by Ishida *et al.* from Hall-effect MR measurements on non-superconducting $BaFe_2As_2$ where we also find that $\mu_{e1} > \mu_h$, although $\mu_{e1}$ is ~5 times smaller and $\mu_h$ is ~58 times smaller in our $SrFe_2As_2$ crystal at 40 K. It is not possible to directly compare the carrier concentrations because they were only listed at 5 K where $n_{e1}$ was significantly less than $n_{e2}$. We also find that $n_{e1}$ in our $SrFe_2As_2$ crystal is small and it increases with decreasing temperature. Ishida *et al.* suggested that the higher mobility and lower carrier concentration of the $e1$ carriers may indicate that it arises from a small electron Fermi surface pocket associated with Dirac-cone-like energy bands [12]. It is possible that $e1$ in our $SrFe_2As_2$ crystal is also associated with electron carriers in a Dirac-cone-like energy band where the Fermi level is above the crossover point. However, if this was true then the absence of quantum linear transport would indicate that the energy difference between the Fermi energy and the crossover point is too large for only one Landau level to be occupied for magnetic fields of up to 8 T.

The magnetic-field dependence of the MR from the 0.8% Ca-implanted crystal (fig. 3(a)) after 72 days in the vacuum desiccator is similar to that seen in the unimplanted crystal after 7 months in a vacuum desiccator (fig. 2(a)) although the MR from the Ca-implanted crystal is lower. There is also a similar crossover in the derivative of the MR as can be seen in fig. 3(b) where the crossover field, $B^*$, is plotted in the inset to fig. 3(a). $B^*$ can also be fitted using eq. 1 where $E_F = 5.5$ meV and $v_F = 2.7 \times 10^5$ ms$^{-1}$. These values are comparable to those found in the unimplanted $SrFe_2As_2$ crystal. However, similar to the MR data from the unimplanted $SrFe_2As_2$ crystal we find that the MR does not have linear magnetic-field dependence in the measured magnetic field range and hence quantum linear transport is unlikely. In fact we find that the 3-carrier transport model provides a good fit to the data as can be seen in fig. 3(a). The fitted carrier mobilities and concentrations are plotted in fig. 3(c) and fig. 3(d). A comparison of the mobilities and carrier concentrations between both crystals can be seen in fig. 3(c) and fig. 3(d) where the fitted data from the unimplanted crystal are also plotted (crosses). We find that $n_{e1} < n_{e2}$ and $\mu_{e1} > \mu_{e2} > \mu_h$ for both crystals. Furthermore, the mobilities are only slightly lower in the 0.8% Ca-implanted crystal (e.g. by up to ~27% at 40 K), which indicates that the low-energy Ca ion implantation has had no significant impact of the carrier scattering. Similarly, Pallecchi *et al.* [9] have found the disorder induced by Ru substitution into $LaFeAsO$ has very weak effect on the band mobilities and transport properties in their quaternary pnictide. The hole and $e2$ carrier concentrations at and above 40 K are also similar but the low $e1$ carrier concentration is lower in the Ca-implanted crystal at and below 60 K. This may suggest that there are some differences in the temperature dependent changes in the $e1$ Fermi surface for both crystals.

The MR from the unimplanted $SrFe_2As_2$ crystal at 20 K is plotted in fig. 4(a) where the MR at 8 T is 197 %. As mentioned earlier the electronic state is inhomogeneous below the emerging $T_c$ of 24 K and hence it is likely that the 20 K MR data arises from suppression of superconductivity by the applied magnetic field in the superconducting regions and 3-carrier transport in the metallic and non-superconducting regions. The suppression of superconductivity can be seen in the low-field part of the MR curve where there is an initial increase in the MR and the MR starts to saturate above ~1 T. This can be clearly seen in the magnetic-field derivative of the MR (inset to fig. 4(a)) where d*MR*/dB rapidly increases below ~1 T. The crossover at ~1 T in the MR is close to the upper critical field, $B_{c2}$ (~ 1 T), measured on a strained-induced superconducting $SrFe_2As_2$ crystal [17]. Furthermore, in fig. 1 it is apparent that at 8 T the resistivity is close to that expected in the normal state and when the MR is dominated by 3-carrier transport. Thus, the MR above ~2 T arises from 3-carrier transport, which was confirmed by fitting the data to eq. (2) where the best fit (red dashed lines) is plotted in fig. 4(a). The fitting was done by assuming that there are no superconducting regions above 2 T. In this case the MR can be written as,

$$MR(B) = MR'(B)/\zeta + (1/\zeta) - 1 \qquad (3)$$

where $MR'(B)$ is the 3-carrier MR in the absence of superconductivity and given by eq. (2), and $\zeta = \rho'(0)/\rho(0)$. $\rho'(0)$ is defined as the resistivity at zero field in the absence of superconductivity. We show in fig. 4(a) that eq. (2) and eq. (3) provide a good fit to the MR where $\zeta$ (20 K) = 0.506. The resultant mobilities are plotted in fig. 2(c) and they are larger than the mobilities are at higher temperatures. The fitted carrier concentrations (fig. 2(d)) are similar or slightly higher than those at higher temperatures.

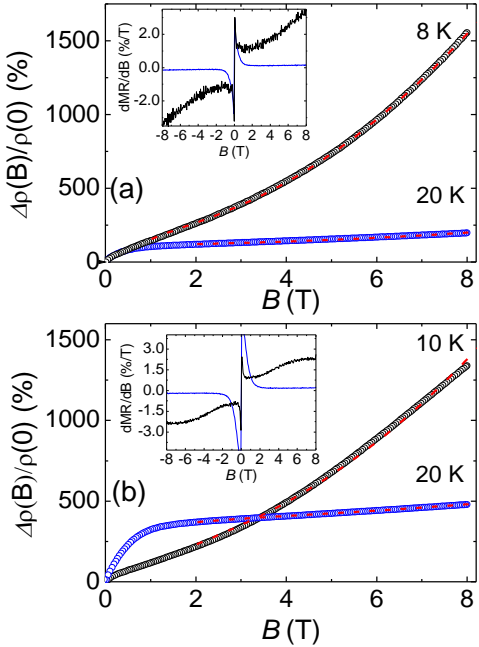

Fig. 4: (Colour on-line) MR for (a) unimplanted SrFe$_2$As$_2$ at 8 K and 20 K after 7 months in a vacuum desiccator, and (b) 0.8% Ca-implanted SrFe$_2$As$_2$ at 10 K and 20 K after 72 days in a vacuum desiccator. The dashed curves are fits to the data assuming a 3-carrier transport model above 1 T or 2 T. The insets show the corresponding derivatives of the MR for the two crystals.

At 8 K the MR increases up to ~0.2 T and then the magnetic-field dependence of the MR changes (fig. 4(a)) and reaches 1600% at 8 T. This change is clearer in the inset to fig. 4(a) where the derivative of the MR is plotted and it can be seen that there are large changes in the derivative below 0.3 T. The initial increase in the MR is indicative of Josephson coupled weak-link behavior that is observed in superconductors when the superconducting regions are badly connected (e.g. in grain-aligned bulk YBa$_2$Cu$_3$O$_x$) [30]. Thus, the change in the magnetic-field dependence of the MR at ~0.2 T suggests that the superconducting regions in the current percolation pathway are badly connected and the weak-links are all driven normal for applied magnetic fields > 0.2 T. The MR at higher applied magnetic fields cannot be due to suppression of superconductivity by the applied magnetic field because the maximum magnetic field of 8 T is far below $B_{c2}$ measured in SrFe$_2$As$_2$ crystals and films [17,18,24]. It is also unlikely that the MR can be attributed to motion of the superconducting vortices that is observed in superconductors and leads to a large increase in the resistivity with applied magnetic field. This is because the irreversibility field at 8 K is close to 8 T in superconducting samples [17,24] and hence flux flow is not expected below ~ 8 T.

The appearance of a finite resistance at 8 K suggests that an appropriate model would be one where there are superconducting droplets in metallic and non-superconducting SrFe$_2$As$_2$. In this case the percolation pathway at 8 K would include conduction in the metallic regions and effectively shorted out superconducting regions. Thus, in a simple model the MR above ~1 T might be expected to be due to the metallic regions and given by eq. (2). For this reason the MR was fitted using eq. (2) and eq. (3) above 1 T. It can be seen in fig. 4(a) that this provides a good fit to the data with $\zeta$ (8 K) = 0.807. We find that the mobilities (fig. 2(c)) are enhanced at 8 K with respect to the values at 40 K and this enhancement is as high as a factor of 8. A large low-temperature enhancement was not reported by Ishida et al. [12] from measurements on non-superconducting BaFe$_2$As$_2$ although the mobilities did increase with deceasing temperature and $\mu_{e1}$ at 8 K is only slightly larger than that reported by Ishida et al. [12]. In our case it is possible that part of the low-temperature increase in the mobilities may be a consequence of the inhomogeneous state and the complex percolation pathway.

The magnetic-field dependences of the MR and the derivative of the MR from the 0.8% Ca-implanted crystal at 10 K and 20 K (fig. 4(b)) are very similar to those from the unimplanted crystal where the MR at 10 K and 8 T reaches ~1300%. In particular, there are similar initial rapid increases in the MR in the low-field region that are either due to the applied magnetic field being above $B_{c2}$ at 20 K or Josephson coupled weak-link behavior at 10 K where the weak-links are driven normal for applied magnetic fields above ~0.3 T. We show in fig. 4(b) that the MR data can also be fitted to 3-carrier transport above 2 T at 10 K and 20 K where the fitted mobilities and carrier concentrations are plotted in fig. 3(c) and fig. 3(d). Similar to the fitted data from the SrFe$_2$As$_2$ unimplanted crystal, we find that the mobilities are enhanced at the lowest measured temperature. However, there is some variation in the absolute values of the mobilities. For example, the hole mobility is larger and the electron mobilities are lower in the 0.8% Ca-implanted crystal when compared with the unimplanted crystal. The electron and hole carrier concentrations are also smaller for the 0.8% Ca-implanted crystal. These differences may be due to a variation in the normal state volume fractions rather than arising from the Ca implantation and it may not be statistically significant if similar measurements were done on a large number of crystals.

**Conclusions.** – In conclusion, we find that the large MR from a pure $SrFe_2As_2$ crystal and a low-energy 0.8% Ca-implanted crystal that were aged by being left in a vacuum desiccator can be fitted to a 3-carrier transport rather than the quantum linear MR model that has been used to model the MR data from some $BaFe_2As_2$ crystals. The MR at low temperatures reaches more than 1600% for the $SrFe_2As_2$ crystal and 1300% for the 0.8% Ca-implanted $SrFe_2As_2$ crystal. The large MR at and below 20 K is most likely due to a 3-carrier transport in an inhomogeneous system with superconducting and metallic regions and where the mobilities are significantly enhanced. There is no evidence that low-energy 0.8% Ca implantation has a significant effect on the carrier mobilities or concentrations. The ability to stabilize this metastable state in $SrFe_2As_2$ by encapsulation will make this iron-pnictide a possible candidate for low-temperature and high-field magnetic sensing applications.

***

**Acknowledgements.** – This work is supported by the Marsden Fund of New Zealand (VUW0917) and The MacDiarmid Institute for Advanced Materials and Nanotechnology. We would like to thank James Storey (VUW) and Uli Zuelicke (VUW) for many useful discussions.

REFERENCES

[1] Kamihara Y., Watanabe T., Hirano M. and Hosono H., *J. Am. Chem. Soc.*, **130** (2008) 3296.
[2] Ishida K., Nakai Y. and Hosono H., *J. Phys. Soc. of Jpn*, **78** (2009) 062001.
[3] Johnston D. C., *Adv. Phys.*, **59** (2010) 803.
[4] Paglione J. and Greene R. L., *Nature Phys.*, **6** (2010) 645.
[5] Lumsden M. D. and Christianson A. D., *J. Phys.: Condens. Matter*, **22** (2010) 203203; Uemura Y. J., *Nature Mater.*, **8** (2009) 253.
[6] Speller S. C., Britton T. B., Hughes G., Lozano-Perez S., Boothroyd A. T., Pomjakushina E., Conder K. and Grovenor C. R. M., *Appl. Phys. Lett.*, **99** (2011) 192504.
[7] Ran Y., Wang F., Zhai H., Vishwanath A. and Lee D. -H., *Phys. Rev. B*, **79** (2009) 014505.
[8] Richard P. *et al.*, *Phys. Rev. Lett.*, **104** (2010) 137001; Richard P., Sato T., Nakayama K., Takahashi T. and Ding H., *Rep. Prog. Phys.*, **74** (2011) 124512.
[9] Pallecchi I. *et al.*, *Phys. Rev. B*, **84** (2011) 134524; Bhoi D., Mandal P., Choudhury P., Pandya S. and Ganesan V., *Appl. Phys. Lett.*, **98** (2011) 172105; Meena R. S. *et al.*, *arXiv*: 1111.3143.
[10] Huynh K. K., Tanabe Y. and Tanigaki K., *Phys. Rev. Lett.*, **106** (2011) 217004.
[11] Tanabe Y., Huynh K. K., Heguri S., Mu G., Urata T., Xu J., Nouchi R., Mitoma N. and Tanigaki K., *Phys. Rev. B*, **84** (2011) 100508(R).
[12] Ishida S. *et. al.*, *Phys. Rev. B*, **84** (2011) 184514.
[13] Kuo H. –H., Chu J. –H., Riggs S. C., Yu L., McMahon P. L., De Greve K., Yamamoto Y., Analytis J. G. and Fisher I. R., *Phys. Rev. B*, **84**, (2011) 054540.
[14] Abrikosov A. A., *Phys. Rev. B*, **58** (1998) 2788; Abrikosov A. A., *Europhys. Lett.*, **49** (2000) 789.
[15] Terashima T. *et al.*, *Phys. Rev. Lett.*, **107** (2011) 176402.
[16] Sutherland M. *et al.*, *Phys. Rev. B*, **84** (2011) 180506(R).
[17] Chong S. V., Tallon J. L., Fang F., Kennedy J., Kadowaki K. and Williams G. V. M., *Europhys. Lett.*, **94** (2011) 37009. doi:10.1209/0295-5075/94/37009
[18] Saha S. R., Butch N. P., Kirshenbaum K., Paglione J. and Zavalij P. Y., *Phys. Rev. Lett.*, **103** (2009) 037005.
[19] Dynamic-TRIM [Biersack J. P., in *Computer simulations of sputtering*, *Nucl. Instrum. and Meth. B*, **27** (1987) 21.
[20] Kasahara S., Hashimoto K., Ikeda H., Terashima T., Matsuda Y. and Shibauchi T., *Phys. Rev. B*, **85** (2012) 060503(R); Rullier-Albenque F., Colson D., Forget A. and Alloul H., *Phys. Rev. Lett.*, **103** (2009) 057001.
[21] Kotegawa H., Sugawara H. and Tou H., *J. Phys. Soc. Jpn.*, **78** (2009) 013709; Colombier E., Bud'ko S. L., Ni N. and Canfield P. C., *Phys. Rev. B*, **79** (2009) 224518.
[22] Lee H., Park E., Park T., Sidorov V. A., Ronning F., Bauer E. D. and Thompson J. D., *Phys. Rev. B*, **80** (2009) 024519.
[23] Ishikawa F., Eguchi N., Kodama M., Fujimaki K., Einaga M., Ohmura A., Nakayama A., Mitsuda A. and Yamada Y., *Phys. Rev. B*, **79** (2009) 172506.
[24] Hiramatsu H., Katase T., Kamiya T., Hirano M. and Hosono H., *Phys. Rev. B*, **80** (2009) 052501.
[25] Chen G. F., Li Z., Dong J., Li G., Hu W. Z., Zhang X. D., Song X. H., Zheng P., Wang N. L. and Luo J. L., *Phys. Rev. B*, **78** (2008) 224512.
[26] Kim J. S., *J. Appl. Phys.*, **84** (1998) 292.
[27] Hu W. Z. *et al.*, *Phys. Rev. Lett.*, **101** (2008) 257005.
[28] Chen G. F. *et al.*, Phys. Rev. B, **78** (2008) 224512.
[29] Ma F., Lu Z. –Y. and Xiang T., Front. Phys. China, **5** (2010) 150.
[30] Wosik J., Xie L. M., Chau R., Samaan A. and Wolfe J. C., *Phys. Rev. B*, **47** (1993) 8968; Hamrita A., Ben Azzouz F., Madani A. and Ben Salem M., *Physica C*, **472** (2012) 34.